\documentclass{aa}

\usepackage{graphicx}
\usepackage[dvipsnames]{xcolor}

\usepackage{txfonts}

\begin{document}

   \title{Galaxy–halo internal alignments across cosmic time}
   
   \author{Agustina V. Marsengo-Colazo\thanks{agustina.colazo@mi.unc.edu.ar}
   \inst{1,2}
          \and
          Facundo Rodriguez\inst{1,2}
          \and
          Manuel Merchán\inst{1,2}
          }

   \institute{CONICET. Instituto de Astronomía Teórica y Experimental (IATE). Laprida 854, Córdoba X5000BGR, Argentina.
       \and
             Universidad Nacional de Córdoba (UNC). Observatorio Astronómico de Córdoba (OAC). Laprida 854, Córdoba X5000BGR, Argentina. }

\abstract
   {Previous observational studies have shown that the principal shape axes of red central galaxies are strongly aligned both with other galaxies in their host group and with the surrounding large-scale cosmic structure. Simulation-based investigations of these intrinsic alignments suggest a link between the alignment of a central galaxy’s major axis with the large-scale galaxy distribution and its internal galaxy–halo shape alignment. In contrast, blue central galaxies typically exhibit little to no alignment signal, owing to a stronger internal misalignment with their halo.}
   {We investigated how the internal alignment between the principal axes of the stellar and dark matter components evolves over time as a function of the total mass of the central galaxy at $z=0$. In particular, we aim to understand why disk-dominated, blue central galaxies often show weak or absent alignment signals with the galaxy distribution in their group and in the larger-scale cosmic structure.} 
   {In this work, we used data from the IllustrisTNG300-1 run and selected a sample of bright central galaxies at $z=0$. From the 3D data, we computed the principal axes of the stellar and dark matter components, along with their angular momenta, to obtain the various alignment angles analyzed in this study. In addition, we used the SUBLINK merger trees to determine the number of major mergers each central galaxy experienced between $z=20$ and $z=0$, and to track their shapes at higher redshifts along their main branch. We examined secondary dependencies of the galaxy-halo alignment on properties such as color and merger history after first dividing the sample into mass bins. Also, we analyzed how shape alignments relate to the dynamical coupling between the angular momentum directions of the stellar and dark matter components.}
   {The results show that central galaxies with  $\mathrm{M_{Tot}} > 10^{13} \, \mathrm{M}_\odot$  tend to align with the shape of their inner halo, largely independent of color or major merger history, although the most massive systems are typically red and have undergone numerous mergers. For lower-mass central galaxies, those that are red and have experienced many mergers exhibit the strongest evolution toward alignment. Blue central galaxies, in contrast, are more strongly influenced by the link between the stellar and dark matter angular momenta, such that they evolve toward either alignment or misalignment with both the shape and angular momentum of the inner halo.}
   {}

   \keywords{dark matter -- Methods: statistical 
   -- Galaxies: halos -- Galaxies: groups: general}

   \maketitle

\section{Introduction}

In the standard cold dark matter  ($\Lambda$CDM) paradigm with a cosmological constant, the matter density  is dominated by dark matter, and initial perturbations in this field grow until sufficiently overdense regions collapse due to gravitational instability, giving rise to dark matter halos. The  halos power the gravitational force that attracts the baryonic content to their potential well, and eventually,  through non-linear and dissipative processes, galaxies  are formed \citep{WhiteRees1978,Blumenthal1984}.
Furthermore, structures develop hierarchically, accreting mass from their surroundings and through mergers with other systems. This leads to tidal effects, environmental influences, and feedback mechanisms that further  complicate this picture.  
In this context, the wide  variety of astrophysical phenomena that galaxies undergo throughout their evolution leave an imprint on the orientations of their shapes and the way they are distributed in space, enhancing or diluting so-called intrinsic alignments, i.e.,  intrinsic correlations between galaxy shape and matter density.  
Therefore, the study of intrinsic alignments can shed light on the complex details of galaxy formation and evolution.

Different kinds of alignments have been measured in observations that  involve the shapes, position angles, and orientations of galaxies on various scales. 
Weak lensing studies use statistical alignment signals as cosmological probes \citep{Kaiser2000,Bacon2000,Takada2004,Munshi2008}. These are apparent alignments in  the observed galaxy shapes  generated by the deviation of  their light  due to the gravitational field of the foreground matter distribution. In this case, intrinsic alignments are a systematic source of error that needs to be corrected for in order to reconstruct the projected mass distribution accurately \citep{Croft2000,Hirata2004,Kirk2012,Krause2016}. Thus, it's currently of utmost importance  to have a precise understanding of intrinsic alignments for the upcoming galaxy surveys that will soon become available and provide better and larger data sets for these studies \citep{Hirata2007,Troxel2015,Joachimi2015,Kirk2015}.

Within groups of galaxies, intrinsic alignments have often been studied  by measuring the probability distribution of the position angles of the satellite galaxies with respect to the  principal axes of the central or brightest group galaxy (BGG).  The evidence shows that satellite galaxies tend to be preferentially located in the direction of the  major axis  of the BGG, and this signal manifests more strongly for red, early-type, and more luminous BGGs. 
Whereas, for  blue  BGGs, the distribution of satellite galaxies seems to be isotropic \citep[see e.g.,][]{SalesyLambas2004,Yang2006,Faltenbacher2007,Agustsson2010,Wang2018,Rodriguez2022}.

To probe intrinsic alignments on larger scales, the anisotropic correlation function, which measures the excess clustering within a certain solid angle in a given direction, has proven to be a valuable tool \citep{Paz2008angular,Paz2011alignments}. This statistic was applied by \cite{Rodriguez2022} on a sample of bright central galaxies identified in the spectroscopic data of the Sloan Digital Sky Survey Data Release 16 \citep[SDSS DR16,][]{Ahumada2020}, where the group finder used is the one described in \citep{Rodriguez2020FOF}. 
This study allowed the detection of an alignment signal between the  position angles of the galaxies and the major axis of the BGGs up to scales greater than $10 \,\mathrm{Mpc}$. Nonetheless, this signal was not present when only blue central galaxies were  taken into account.  \cite{Rodriguez2023} analyzed further the alignment of bright central galaxies in the IllustrisTNG hydrodynamical simulation at $z=0$ and concluded that the anisotropy was due to a series of alignments linked at various scales. First, the baryonic component of the central galaxy aligns  with the shape of its own halo; then the halo aligns with the host halo of the group which, in turn, tends to be aligned with the surrounding matter. Also, they found that the lack of anisotropy for blue centrals is due to a misalignment between their stellar component and the shape of their halo.  
These results were complemented in \cite{Rodriguez2024}, where the same sample of central galaxies in  IllustrisTNG was used to study the evolution of the anisotropic correlation function  using the principal axes of both the stellar and dark matter shapes. The results yielded an increasing alignment, as redshift decreases, of the galaxy distribution with the major axis of the stellar component, but once again, only for red centrals. 

Many works in simulations also show that intrinsic alignments depend on redshift, as well as on color and morphology type. 
For example, \cite{Debattista2015} simulated galactic disks in individual triaxial halos with different initial orientations relative to the halo's principal axes, and studied their evolution over time. The authors concluded that, in most configurations, the  red disks without gas tend to align their spin  with the short axis of a triaxial halo as they evolve. The angular momentum of blue disks, on the other hand, is affected by the infalling cool gas that may allow it to be more misaligned with the halo. 
Following another approach, \cite{Xu2023}  used the IllustrisTNG300-1 hydrodynamical simulation  to study  the mean alignment of central galaxies between the principal axes of the stellar component within twice the stellar half-mass radius and the dark matter contained in the central subhalo, from $z=1.5$ up to $z=0$. Their results show that the alignment is stronger for those galaxies that have a higher fraction of accreted mass from galaxy mergers $F_{acc}$,  a larger halo and stellar mass. Whereas, the alignment decreases with redshift and for late-type galaxies. Also, they conclude that $F_{acc}$ seems to be the parameter that more fundamentally determines the galaxy-halo alignment, in the sense that when that parameter is  fixed, the mass and redshift dependence are  significantly reduced.

Going beyond these internal alignments, it has been established in different hydrodynamical simulations that within the host halo, satellite galaxies are typically distributed on the galactic plane of the central galaxy and aligned with the major axis. This alignment signal is also stronger for red, passive, and spheroidal galaxies, and it increases with halo mass   \citep[e.g.,][]{Dong2014,Welker2018,Tenneti2021,Zjupa2022}. However, some differences have arisen in the  redshift dependence.
\citet{Tenneti2014} studied galaxy–halo alignments in the cosmological MassiveBlack-II simulation and found only a weak dependence of the alignment signal on redshift between $z=1$ and $z=0.06$. In contrast, \citet{Chisari2017} showed that in the Horizon-AGN simulation, galaxy–halo alignments tend to increase from $z=3$ to $z=0$. These discrepancies may arise from the use of different simulations, as baryonic processes are implemented through distinct subgrid models that can affect stellar shape orientations. Although simulations inherently carry this type of uncertainty, they remain valuable tools for assessing the qualitative behavior of intrinsic alignments.

Alignments are commonly thought to be driven by tidal effects; therefore, the angular momentum of galaxies is an important factor to take into account. 
In particular, the  orientation of late-type, blue galaxies  is closely related to the direction of their angular momentum, given that their spin axis  defines the plane of their disk, and this spin is thought to be closely related to the angular momentum of the galaxy's halo \citep[e.g.,][]{Bailin2005}. 
The origin of the galaxies' angular momentum has been a historic topic of importance. Currently, the most accepted explanation is given by tidal torque theory, which proposes that the halos acquired their angular momentum through the external torques exerted by the large-scale matter distribution acting on non-spherical overdensities that are initially misaligned with the tidal field \citep[e.g.,][]{Hoyle1951,Doroshkevich1970,White1984,HeavensyPeacock1988,Catelan1996,LeeyPen2000,Codis2015}. 
Furthermore, the baryons at first share the same specific angular momentum distribution as the dark matter, since in the linear regime both components are dynamically coupled \citep{Fall1980}. 
However, as it was stated above, the galaxy and dark matter angular momenta  may not remain aligned as the baryons are subjected to further non-linear, astrophysical processes and  perturbations. 
In fact, studies in hydrodynamical simulations have found a median misalignment of $\sim 30^\circ$   between the angular momenta of the baryons and the dark matter halo of disk galaxies  \citep[e.g.,][]{VandenBosch2002,Yoshida2003,Chen2003,Bett2010,Zjupa2022}. 
Therefore, given the results of \cite{Debattista2015} showing that the inflow of gas with misaligned angular momentum helps to keep the disk minor axis misaligned relative to the halo’s minor axis, there should be a connection between the galaxy–halo shape alignment in blue late-type galaxies and their galaxy–halo angular momentum alignment. Moreover, the authors found that as long as the angular momentum of the gas changes orientation, the spin direction of the disk must also vary to remain in equilibrium. This provides further evidence that even isolated disk galaxies can undergo significant changes in orientation over time.

In addition, the condensation of  baryons in the halo's potential well is expected to affect the inner  halo's triaxial shape, well beyond the size of the galaxy. Dynamically, the predominant type of orbits that sustain triaxial mass configurations are box orbits that cross the center of the system. When a baryonic disk is formed in the triaxial halo,  the radial structure of the dark matter is modified and the central density increases. This gives rise to the so-called adiabatic contraction effect, but most importantly, box orbits are deformed, making the halo rounder and nearly axisymmetric  \citep[e.g][]{Dubinski1994,Gnedin2004,Debattista2008,Kazantzidis2010,Chua2019}. 
Although the halo’s exact response also depends on the specific manner in which the baryons are assembled, that is, even when the same amount of baryonic mass is considered, the halo’s response differs if the mass is smoothly accreted or driven by mergers \citep{Abadi2010}.
Therefore, a better understanding of  the coupled evolution of the stellar shape and the not directly observable dark matter may provide better constraints for the dynamical modeling of dark matter halos and reveal new clues about the nature of dark matter itself. 
On the other hand, the outer halo's shape seems to be less affected by the inner galaxy and more influenced by the environment and large scale anisotropies of  infalling material \citep[e.g.,][]{Bailin2005,Valenzuela2024,Han2024,Rodriguez2025}.
Therefore, the connection between the internal and large scale alignments reflects the interplay  between these multiscale processes.

Hence, in this paper, we focus on studying the evolution of the internal alignment between the principal axes of the stellar component and those of the inner halo, as well as its dependence on mass, using data from the IllustrisTNG300 hydrodynamical simulation. To shed light on the differences in alignment between red and blue populations reported by \cite{Rodriguez2023,Rodriguez2024}. We further explore secondary dependencies on other properties, such as color and merger history, after binning the sample by mass.  
Finally, we  examine the relationship between shape alignments and the dynamical coupling of the stellar and dark matter angular momentum directions. 
This analysis aims to provide insight into why disk-dominated, blue central galaxies often lack a clear alignment signal with the distribution of galaxies in their group as well as on larger scales.

This paper is organized as follows. Section \ref{sec:data} describes the simulation data used in this study and the methodology for computing the relevant properties, shapes, and alignment angles. Section \ref{sec:galaxy-halo shape} presents the evolution of stellar–dark matter shape alignments as a function of mass and examines their dependence on color and major mergers. Section \ref{sec:angular momentum} focuses on angular momentum–related alignments and their connection to the shape alignments of lower-mass BGGs. Finally, Section \ref{sec:conclusions} summarizes the main findings and conclusions of this work.

\section{IllustrisTNG simulation data}
\label{sec:data}

In this study, we use data from the IllustrisTNG300-1 run (hereafter TNG300-1) of the IllustrisTNG project \citep{Weinberger2017, Springel2018, Naiman2018, Marinacci2018, Pillepich2018, Nelson2018, Nelson2019}. The IllustrisTNG suite is a set of magnetohydrodynamic cosmological simulations that employ the AREPO moving-mesh code \citep{Springel2010_MovingMesh}, following the standard $\Lambda \mathrm{CDM}$ cosmology with parameters $\Omega_{\rm m} = 0.3089$,  $\Omega_{\rm b} = 0.0486$, $\Omega_\Lambda = 0.6911$, $H_0 = 100\,h\, {\rm km\, s^{-1}Mpc^{-1}}$ with $h=0.6774$, $\sigma_8 = 0.8159$, and $n_s = 0.9667$ \citep{Planck2016}. The simulations apply subgrid models that include a wide range of physical processes, including star formation, chemical enrichment of the interstellar medium from Type II and Ia supernovae, as well as AGB stars, stellar and AGN feedback, radiative cooling influenced by metal content, among others \citep{Vogelsberger2013}.
In particular, the TNG300-1 run is a simulated box with a side length of $205 \, h^{-1} \, \mathrm{Mpc}$, representing the highest-resolution run within the TNG300 suite, populated with $2500^3$ dark matter particles of mass $4.0 \times 10^7\,h^{-1}\,{\rm M_{\odot}}$  and $2500^3$ gas cells of mass $7.6 \times 10^6\,h^{-1}\,{\rm M_{\odot}}$. 
The outputs are registered in 100 snapshots spanning redshifts from $z = 20$ to $z = 0$.

The dark matter halos at each snapshot were identified with a Friends of Friends (FOF) algorithm with a linking length of 0.2 times the mean interparticle separation \citep{Davis1985}. We will refer to these host halos as groups  throughout this paper. On the other hand, the SUBFIND algorithm identifies gravitationally bound substructures within each halo, the so-called subhalos  \citep{Springel2001_fof,Dolag2009_fof}. 
As previously mentioned, \cite{Rodriguez2024} analyzed a sample of bright BGGs selected at $z = 0$ in the TNG300-1 simulation, revealing differences in the evolution of the anisotropic correlation function between the red and blue populations. In the present work, we investigate the evolution of the internal alignment for the same sample. To this end, we use the snapshots $99,\,92,\,86,\,78,\,71,\,63,\,56,\,48,\,40$ and $33$ corresponding to $z=0,\, 0.08,\, 0.17,\, 0.3,\, 0.42,\, 0.6,\, 0.8,\, 1.07,\, 1.5$ and $2 $, which were selected to be approximately evenly spaced in the scale factor $a$.

First, we selected all simulated galaxies at $z = 0$ with stellar masses greater than $10^{8.5} \, \mathrm{M}_{\odot}$ to avoid resolution issues. Our analysis focuses specifically on central galaxies, defined as the most massive galaxy of each group,and they correspond to those subhalos whose index matches the GroupFirstSub field of their associated FoF group in the TNG300-1 simulation.  To connect the results of this work back to the large-scale alignments studied in \cite{Rodriguez2023,Rodriguez2024}, we followed their selection criteria  and further required that the BGGs' absolute magnitude in the r-band satisfy $\mathrm{M_r} < -21.5$, resulting in a total of $14135$ BGGs.
From the halo and subhalo catalogs at $z = 0$ of the TNG300-1 simulation, we extracted several properties, including the stellar, gas, and dark matter masses of each galaxy, in order to calculate the total mass $\mathrm{M}_\mathrm{Tot}$, defined as the sum of the mass of each component, and the absolute magnitude in  $g$ and $r$  bands. Using the SUBLINK merger trees \citep{Rodriguez-Gomez2015}, we then traced back this population of subhalos to earlier redshifts.

\subsection{Shape and angular momentum angles} 

We assume that the shape of galaxies and  subhalos can be modeled as triaxial ellipsoids. In this framework, the eigenvalues and eigenvectors of the inertia tensor provide the lengths and orientations of the principal axes of each structure. 
To minimize the effect of particles at large radii that could induce asymmetries in the shape calculation \citep{Zemp2011, Bassett2019}, we compute the stellar (dark matter) inertia tensor using stellar (dark matter) particles within a sphere of radius twice the stellar (dark matter)  half-mass radius.
The components of the inertia tensor  are calculated as
\begin{equation}
    I_{ij} = \sum_n m_n x_{i,n} x_{j,n}\,,
\end{equation} 
where $i$ and $j$ refer to the  spatial axes of the simulated box (i.e., \(i, j = 1, 2, 3\)), and  \(m_n\) is the mass of the \(n\)-th particle. The quantities \(x_{i,n}\) and \(x_{j,n}\) denote the components of the position vector of the \(n\)-th particle along the \(i\)-th and \(j\)-th axes, respectively. The position vectors are measured relative to the subhalo's center, defined as the location of the particle with the minimum gravitational potential energy.  
Then, by diagonalizing the inertia tensor of both the stellar and dark matter component, the eigenvectors  that correspond to the largest, intermediate and smallest eigenvalues $\lambda_a,\, \lambda_b,$ and $\lambda_c$, respectively, provide the orientations of the major, intermediate and minor axes of each structure. Also, the length of each axis is given by the square root of each eigenvalue, i.e., $\sqrt{\lambda_a}=a$, $\sqrt{\lambda_b}=b$ and $\sqrt{\lambda_c}=c$ , where $a>b>c$.
With this information, we quantify the internal shape alignments using the angles $\theta_a$ and $\theta_c$, defined as
\begin{equation}
%\centering
\cos(\theta_a) = \frac{ | \mathbf{a}_{*} \cdot \mathbf{a}_\mathrm{DM} |  }{|\mathbf{a}_{*}|\,|\mathbf{a}_\mathrm{DM}|} 
\end{equation}
and
\begin{equation}
%\centering
\cos(\theta_c) = \frac{ | \mathbf{c}_{*} \cdot \mathbf{c}_\mathrm{DM} |  }{|\mathbf{c}_{*}|\,|\mathbf{c}_\mathrm{DM}|} \, ,
\end{equation}
where $\mathbf{a}_\mathrm{*}$ and  $\mathbf{c}_\mathrm{*}$ are the major and minor axes of the galaxy, while $\mathbf{a}_\mathrm{DM}$ and  $\mathbf{c}_\mathrm{DM}$ are the major and minor axes of the inner halo. 
If the galaxy and subhalo shapes were  uncorrelated, the orientations of their semi-axes would be random. 
In that case, the average angles $\langle \theta_a \rangle$ and $\langle \theta_c \rangle$ would be $60^\circ$. Therefore, average values greater than $60^\circ$ indicate misalignment between the semi-axes, while values smaller than $60^\circ$ suggest a tendency toward alignment.

Similarly, we calculate the galaxy (dark matter) angular momentum as the sum of the angular momentum of all stellar (dark matter) particles within twice the stellar (dark matter) half-mass radius:
\begin{equation}
    \vec{\mathrm{J}} = \sum_n m_n  (\vec{r}_n \times \vec{v}_n) \, ,
\end{equation} 
where \(m_n\) is the mass of the \(n\)-th particle, while \(\vec{r}_{n}\) and \(\vec{v}_{n}\) are the position and velocity vectors of each particle measured relative to the subhalo's center. 
In this work, we focus solely on the orientation of the angular momentum. 
To this end, we compute the angle between the galaxy's minor axis and its angular momentum 
\( \theta_{cJ} \), calculated as 
\begin{equation}
%\centering
\cos(\theta_{cJ}) = \frac{ | \mathbf{c_{*}} \cdot \mathbf{J_{*}} |}{|\mathbf{c_{*}}|\,|\mathbf{J_{*}}|} \, ,
\end{equation}
where  $\mathbf{J}_{*}$ is the stellar angular momentum.

Also, we are interested in the alignment between the angular momentum of the galaxy and that of its subhalo, quantified as
\begin{equation}
\cos(\theta_{J}) = \frac{ \mathbf{J}_{*} \cdot \mathbf{J}_{\mathrm{DM}} }{|\mathbf{J}_{*}|\,|\mathbf{J}_{\mathrm{DM}}|} \, ,
\end{equation}
where $\mathbf{J}_{\mathrm{DM}}$ is the  dark matter angular momentum vector.
Notice that, for this angle, we don't take the absolute value of the numerator because, in this case, parallel and antiparallel angles represent two distinct physical configurations. Thus, \( \theta_{J} \) ranges from $0$ to $180^\circ$, and for two random directions, the average value is $\langle \theta_J \rangle = 90^\circ $.

\subsection{Merger trees}

To study the evolution of alignments, we employed the subhalo merger trees constructed with the SUBLINK algorithm \citep{Rodriguez-Gomez2015}, and followed the main progenitor branch of each BGG selected at $z = 0$. This main branch corresponds to the sequence of progenitors with the most massive history, as defined in \cite{DeLucia2007_MainProg}, providing the ID of the main progenitor at each redshift. Using the stellar and dark matter particles associated with these main progenitors, we compute the stellar and subhalo shapes and their alignment for the different snapshots already mentioned. 

Also, to explore the effect of  major mergers on the internal alignments,  we use the full merger tree history of the BGGs. 
First, we identify, within the data structure of the merger tree, the subhalos that  merged with each BGG throughout its entire history, that is, from snapshot $0$ (corresponding to $z=20$) to snapshot $99$ (corresponding to $z=0$). A merger is considered to occur when two subhalos in the tree are associated with the same descendant subhalo. Thus, if a subhalo has $\mathrm{N}_p$ progenitors, it means that its main progenitor experienced $\mathrm{N}_p -1$ mergers \citep{Rodriguez-Gomez2015}.
The progenitors of the resulting subhalo are generally associated with the previous snapshot. However, since the SUBLINK algorithm allows subhalos to skip a snapshot, it is possible that some progenitors are located two snapshots earlier.
Finally, using this information, the  number of major mergers was calculated for each central galaxy by counting the total number of mergers characterized by a stellar mass ratio  greater than $0.2$.

\section{Evolution of shape alignments}
\label{sec:galaxy-halo shape}

Previous studies have shown that the existence of  preferred directions in the large-scale galaxy distribution along the principal axes of BGGs is connected to how strongly the BGGs' stellar shape is  aligned with its dark matter halo. These works also found that the red and blue galaxies exhibit distinct alignment behaviors \cite{Rodriguez2023,Rodriguez2024}. 
\begin{figure}[ht!]
    \centering
    \includegraphics[width=0.4\textwidth]{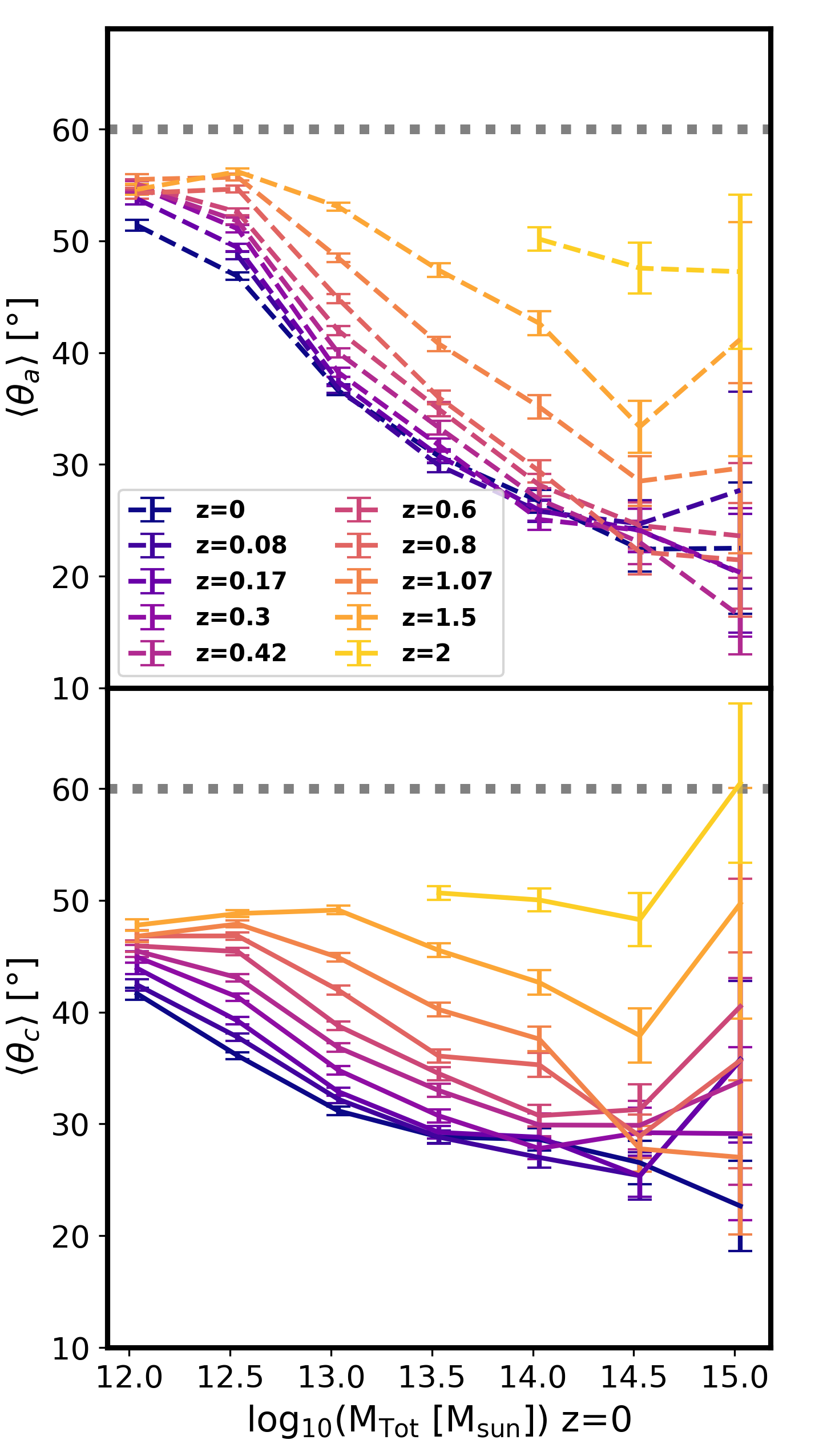}
    \caption{Average misalignment angles  $\theta_a$ (upper panel) and $\theta_c$ (lower panel) of the  stellar and dark matter major and minor axes respectively calculated at redshifts $z=0,0.8,0.17,0.3,0.42,0.6,1.07,1.5,2$, as a function of the total mass (i.e., the sum of the stellar, gas and halo masses) at $z=0$. Lighter colors correspond to higher redshifts. The grey dotted line indicates an average angle of $60^{\circ}$, corresponding to random orientation, smaller angles indicate a tendency toward alignment. Error bars are calculated using the standard deviation of the mean.} 
    \label{aStyaDm(masa)_Total}
\end{figure}
\begin{figure*}
\sidecaption
  \includegraphics[width=12cm]{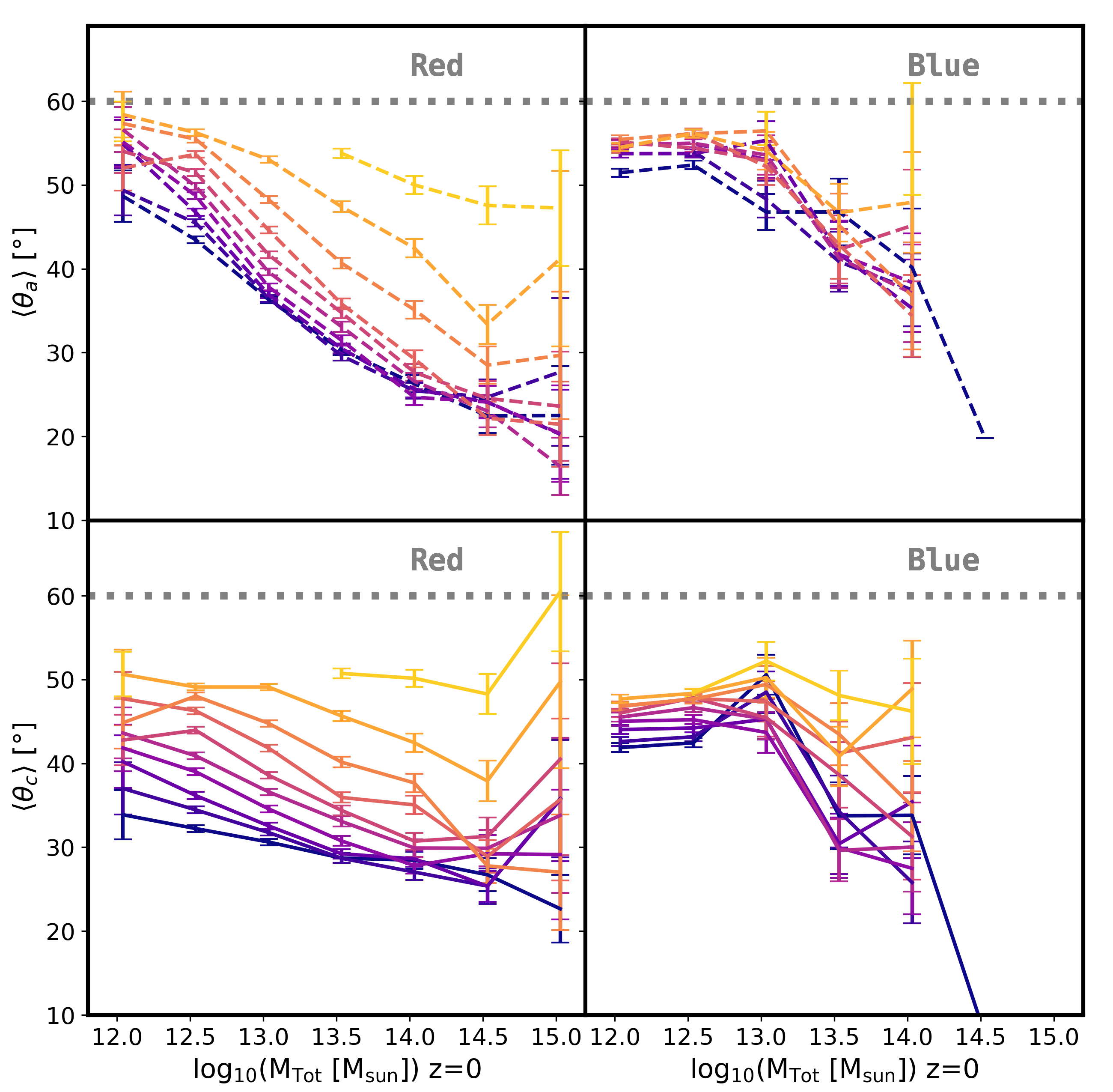}
     \caption{Evolution of the average misalignment angles  $\theta_a$ (upper panels) and $\theta_c$ (lower panels) of the  stellar and dark matter major and minor axes respectively, as a function of total mass. Left panels show the evolution of the alignments of  BGGs that are classified as red at $z=0$, i.e $(g-r)_{z=0}>0.6$, while the right panels show the evolution of  blue BGGs  i.e $(g-r)_{z=0}<0.6$.}
     \label{aStyaDm(masa)_color}
\end{figure*}
Thus, we want to better characterize the evolution of  the angles between the principal axes of the stellar and dark matter shapes within twice their respective half-mass radii, hereafter referred to as internal alignments. For this, we use the same sample of simulated galaxies from the TNG300-1 run as in those  previous works. 
Also,  it has been shown that alignments present a significant dependence on mass, and this is arguably one of the most fundamental  properties of galaxies, so our approach is to first bin the sample by mass and then investigate secondary dependencies on other properties  \citep[e.g.,][]{Xu2023,Rodriguez2025}.

Therefore, we focus first on  the evolution of the internal  alignments between the major and minor axes of the stellar and dark matter components as a function of the total mass fixed at $z=0$.  
The upper and lower panels of Fig. \ref{aStyaDm(masa)_Total} show that the angles $\theta_a$ and $\theta_c$, respectively, decrease with increasing mass at each redshift, although the trend is strongest at $z=0$, meaning that galaxies with higher total mass at $z=0$ trace better the internal shape of their halo. 
However, these massive galaxies did not always present such alignment; rather, they evolved over time, with their stellar and dark matter shapes becoming increasingly aligned as redshift decreases. In contrast, lower-mass galaxies experienced less evolution in this regard, primarily improving their alignment along the minor axis but to a lesser extent. 
The greater alignment signal in $\theta_c$ than in $\theta_a$ for BGGs with lower mass is probably due to the fact that these are predominantly late-type and, therefore,  oblate in shape (see Fig. 12 in \cite{Rodriguez2024}). This implies that the minor axis direction is well defined, while the major and intermediate axes are similar in magnitude and lie within the plane defined by the galactic disk, but they are indistinguishable in orientation due to their degeneracy. Therefore, from now on, when analyzing the alignment of  lower-mass galaxies, we will focus on the behavior of their minor axis and only show their alignment along the major axis for completeness.

\begin{figure*}[ht!]
    \centering
    \includegraphics[width=0.3\textwidth]{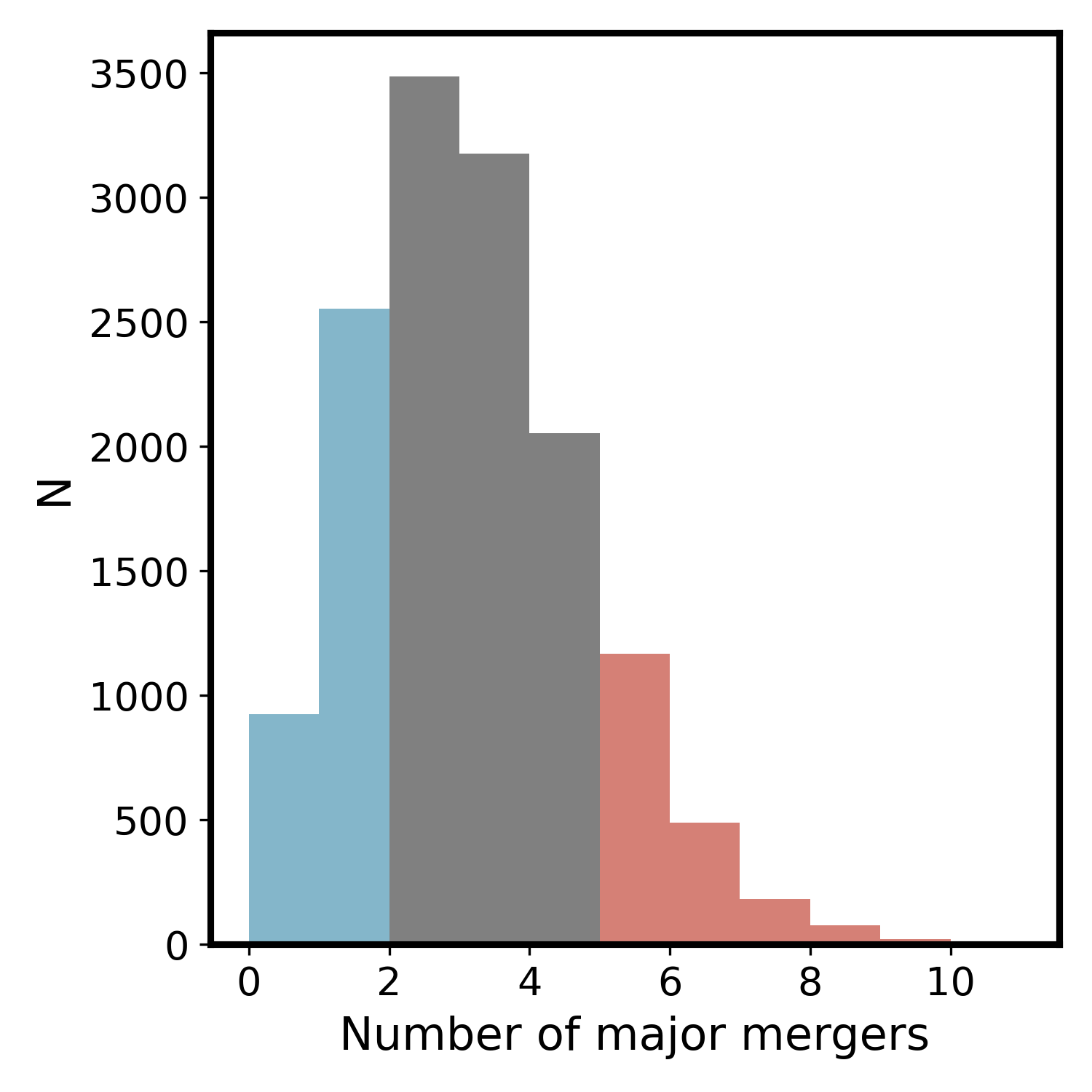}
    \includegraphics[width=0.3\textwidth]{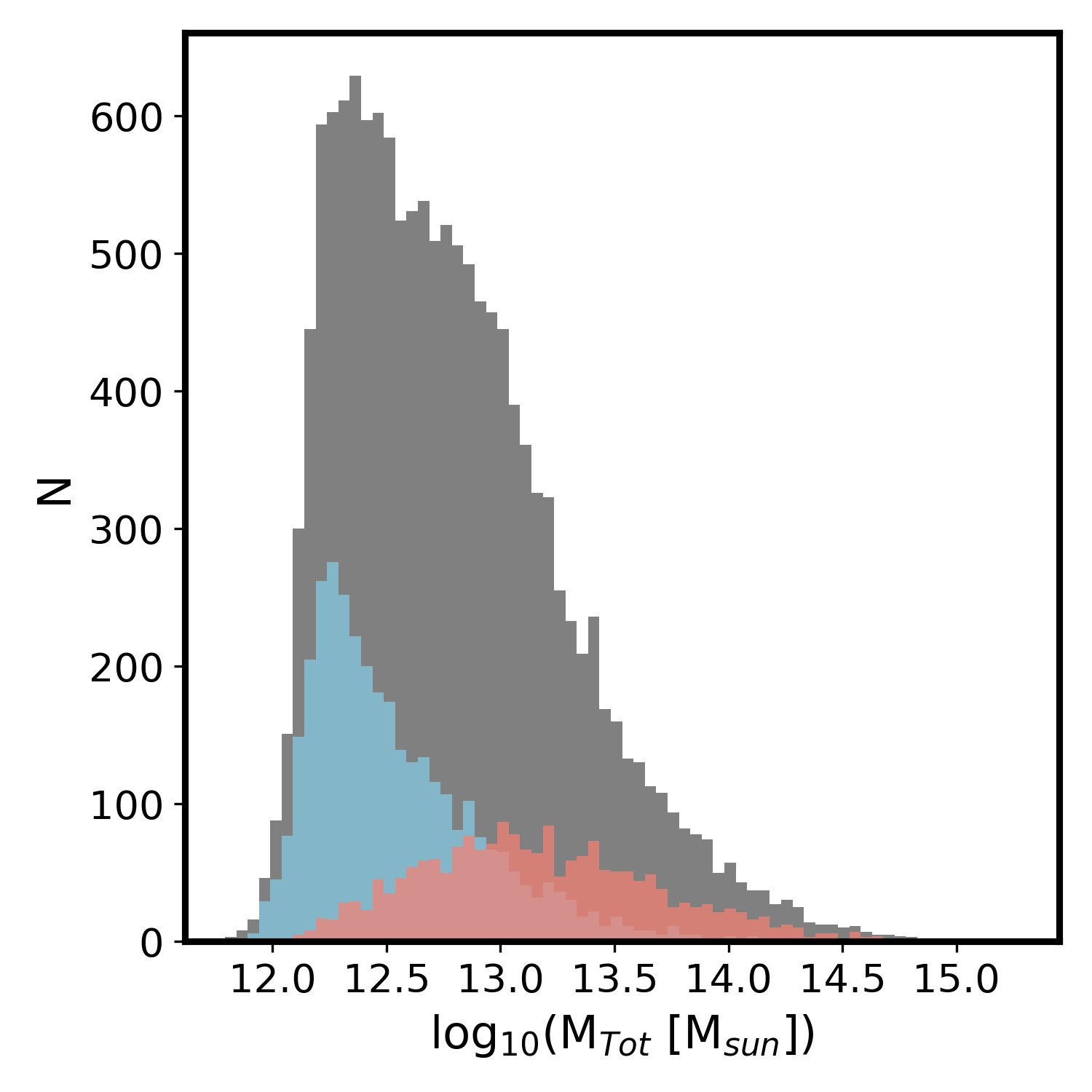}
    \includegraphics[width=0.3\textwidth]{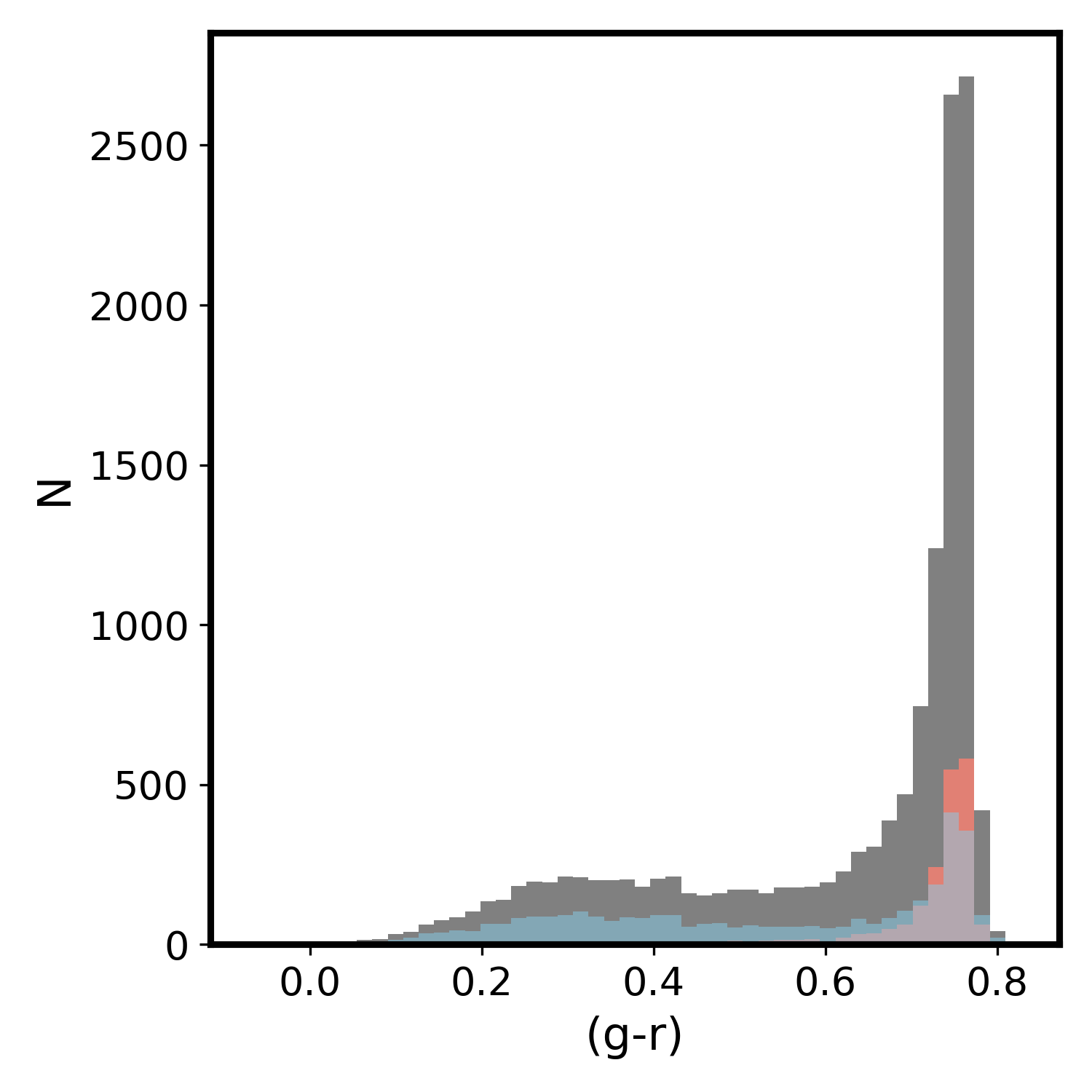}
    \caption{Left, middle, and right panels illustrate the histogram of the number of major mergers from $z=15$ to $z=0$, total mass and color, respectively, for the entire sample of BGGs shown in grey. In light blue we select the galaxies that had 1 or 0 major mergers and in pink those that had 5 or more major mergers.} 
    \label{Hist_Nmerg}
\end{figure*}
\begin{figure*}
\sidecaption
  \includegraphics[width=12cm]{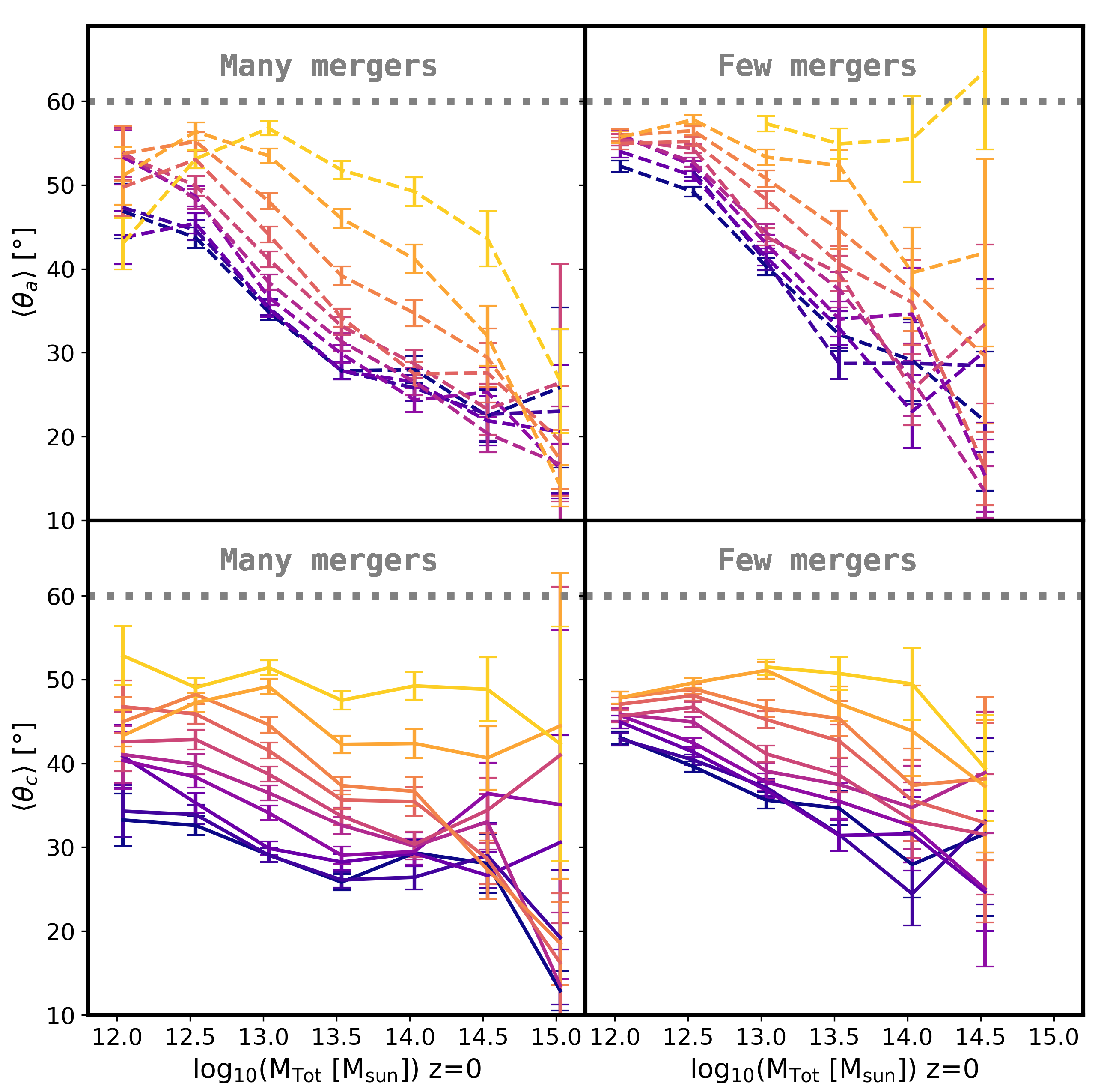}
     \caption{Evolution of the average misalignment angles  $\theta_a$ (upper panels) and $\theta_c$ (lower panels) between the  stellar and dark matter major and minor axes respectively, as a function of total mass. Left panels show the evolution of the alignments for BGGs that have experienced many mergers ($ \mathrm{N}_\mathrm{Merg} \ge 5$) from $z=15$ to $z=0$, while the right panels show the evolution of BGGs with few mergers ($ \mathrm{N}_\mathrm{Merg} \le 1$).}
     \label{aStyaDm(masa)_Nmerg}
\end{figure*}
\begin{figure*}
\sidecaption
  \includegraphics[width=12cm]{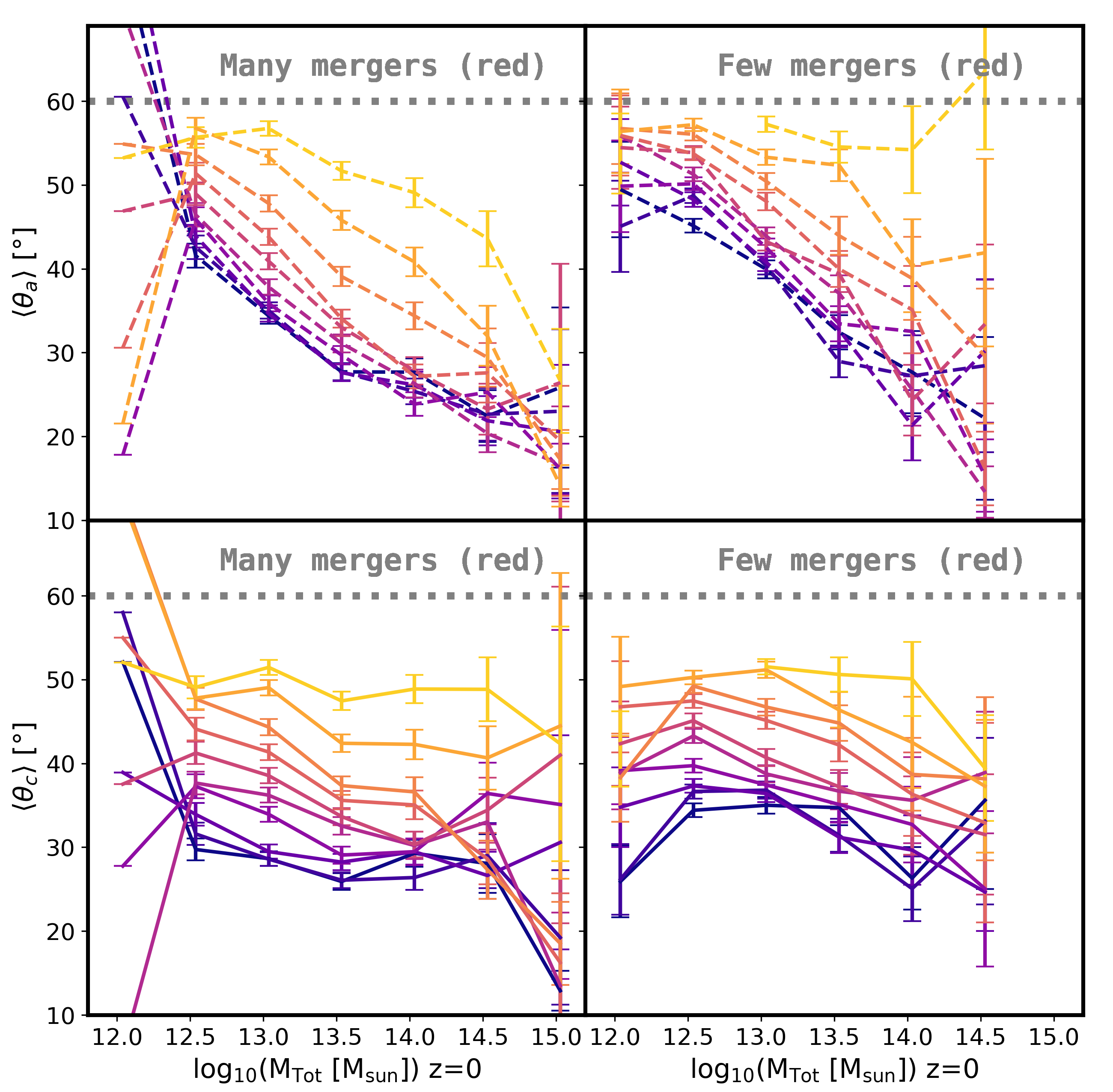}
     \caption{Evolution of the average misalignment angles  $\theta_a$ (upper panels) and $\theta_c$ (lower panels) between the  stellar and dark matter major and minor axes respectively, as a function of total mass. Left panels show the evolution of the alignments for red BGGs that have experienced many mergers ($\mathrm{N}_\mathrm{Merg} \ge 5$) from $z=15$ to $z=0$, while the right panels show the evolution of red BGGs with few mergers ($\mathrm{N}_\mathrm{Merg} \le 1$).}
     \label{aStyaDm(masa)_NmergyRed}
\end{figure*}

\subsection{Color}

In order to explain the differences in the alignment signal for the red and blue BGGs in \cite{Rodriguez2023,Rodriguez2024},  we now analyze  the evolution of the internal alignments for these two populations. For this, we select red and blue BGGs at $z=0$ as those that satisfy that $(g-r)>0.6$ and  $(g-r)<0.6$, respectively, using the same criteria as in \cite{Rodriguez2023,Rodriguez2024},  and follow  their alignments back in time. 
The left  and right panels of Fig. \ref{aStyaDm(masa)_color}  show the evolution of red and blue BGGs, respectively.  In both cases, $\theta_a$ and $\theta_c$ decrease with mass and redshift, but red BGGs dominate the alignment behavior for the higher mass systems, exhibiting a remarkable evolution towards alignment with the dark matter. Also, the alignment in $\theta_c$ of red BGGs at the low mass end is  stronger than that of the total sample. 

Blue BGGs at high redshift show a similar alignment signal as the red population. However, for blue central galaxies with lower masses ($\mathrm{M}_\mathrm{Tot} \leq 10^{13} \, \mathrm{M}_\odot$), it is noticeable that $\theta_c$ decreases only mildly, resulting in a weaker alignment at $z=0$. High-mass blue BGGs, on the other hand, tend to increase their alignment $\theta_c$ more strongly, although the trend is noisier because of the fewer number of blue galaxies in that mass range. Therefore, this suggests that blue BGGs are not as coupled to their internal dark matter shape.

\subsection{Major mergers}

Mergers may play a twofold role: on the one hand, they are expected to trace anisotropies in the large-scale matter distribution and thus enhance large-scale alignments; on the other hand, they can disrupt stellar orbits and modify galaxy shapes, leading to a dilution of pre-existing internal alignments.
Therefore, we examined how the internal alignment is affected by the total number of mergers experienced by BGGs from $z=15$ to $z=0$, considering only those events in which the satellites contributed at least $20\%$ of the BGG's stellar mass.

In the left panel of Fig. \ref{Hist_Nmerg}, we show the distribution of the number of major mergers for the full sample of central galaxies. We note that the minimum is $0$ (i.e.,  BGGs that experienced no major mergers), while the maximum observed value is $10$. BGGs that have experienced five or more major mergers over their evolutionary history are shown in pink, while those that have had zero or one major merger are shown in light blue. Overall, it seems to be relatively rare for the galaxies in this sample to undergo many major mergers,  as the median of the distribution is three major mergers.  

Next, we inspect in the middle panel of Fig. \ref{Hist_Nmerg} the mass distribution for the population with few mergers (light blue) and many mergers (pink). We observe that central galaxies with few major mergers are typically on the low mass end, which makes sense given that they probably inhabit lower density environments where mergers are not very frequent, so they must have acquired the bulk of their mass through smooth matter accretion from their surroundings. Meanwhile, galaxies with many major mergers have a rather broad mass distribution that peaks at $\sim 10^{13} \, \mathrm{M}_\odot $. 
On the other hand, evaluating the color distribution  of these two populations in the right panel of Fig. \ref{Hist_Nmerg}, we note that BGGs with many mergers are red, whereas those that have experienced few mergers have a broad color distribution, mixing both red and blue galaxies. This implies that those BGGs that had many mergers are quenched at $z=0$, but those that only experienced at most one major merger in some cases are star forming, while others may have been quenched by  other processes.

To study the influence of the number of major mergers, we analyze in Fig. \ref{aStyaDm(masa)_Nmerg} the evolution of $\theta_a$ and $\theta_c$ for BGGs that have experienced many mergers (left panels) and few mergers (right panels). 
Overall, we find that at the high-mass end, the major and minor axes of BGGs become increasingly aligned as redshift decreases, with only a mild dependence on the number of major mergers. Notably, BGGs that experienced numerous mergers already show enhanced alignment in their major axis angle ($\theta_a$) at high redshift, whereas those with fewer mergers display an almost random orientation at $z=2$ and gradually align with the dark matter major axis over time.
In contrast, at the low-mass end ($\mathrm{M}_\mathrm{Tot} \leq 10^{13} \, \mathrm{M}_\odot$), the misalignment angles $\theta_a$ and $\theta_c$ decrease more markedly for the population with many mergers compared to those with fewer mergers in the same mass range. In particular, the evolution of the shape alignment for BGGs with multiple mergers resembles that observed for the red population.

These results suggest that as galaxies grow in mass, their major-axis alignment strengthens more rapidly when driven by mergers, whereas if the mass is predominantly accreted smoothly, the alignment develops more gradually.
For lower-mass galaxies, major mergers also seem to enhance their internal alignment, although given that  BGGs that have experienced many mergers tend to be red, as shown in \ref{Hist_Nmerg}, it is hard to say which parameter is driving the alignment in this case.

Thus,  to disentangle the effects of quenching and major mergers on the alignments, in Fig. \ref{aStyaDm(masa)_NmergyRed} we show the evolution of  $\theta_a$ and $\theta_c$ after selecting only red BGGs that have experienced many major mergers (left panels) and few major mergers (right panels). 
Now that both populations are not affected by color, the differences highlighted above for the lower mass BGGs get diluted, and naturally the trends appear noisier due to the smaller number of galaxies considered. Nonetheless, red galaxies with many major mergers still have slightly greater internal alignments both in $\theta_a$ and $\theta_c$. This suggests that color has a strong correlation with shape alignments, once the mass is fixed, and that the number of major mergers induces secondary effects. In consequence, when galaxies lose their gas supply and star formation ceases, the stellar component tends to align over time with the shape of the dark matter halo. Since major mergers are among the processes responsible for quenching, this contributes to a stronger alignment signal in the population of BGGs that have experienced many such events. However, major mergers themselves appear to promote the alignment of the stellar shape with that of the inner halo beyond the reddening effect alone. Therefore, a more detailed analysis focused on other merger-related properties would be valuable for future work, in order to better understand their role in shaping these alignments.

\section{Angular momentum alignments}
\label{sec:angular momentum}

The orientation of blue late-type BGGs is thought to be closely linked to their angular momentum. Hence, to further investigate why the alignment of the minor axis of blue BGGs with the dark matter minor axis is weaker than that observed for red BGGs, particularly at the low-mass end, we examine in this section the evolution of the stellar angular momentum direction and its connection to that of the halo.

\begin{figure}[ht!]
    \centering
    \includegraphics[width=0.4\textwidth]{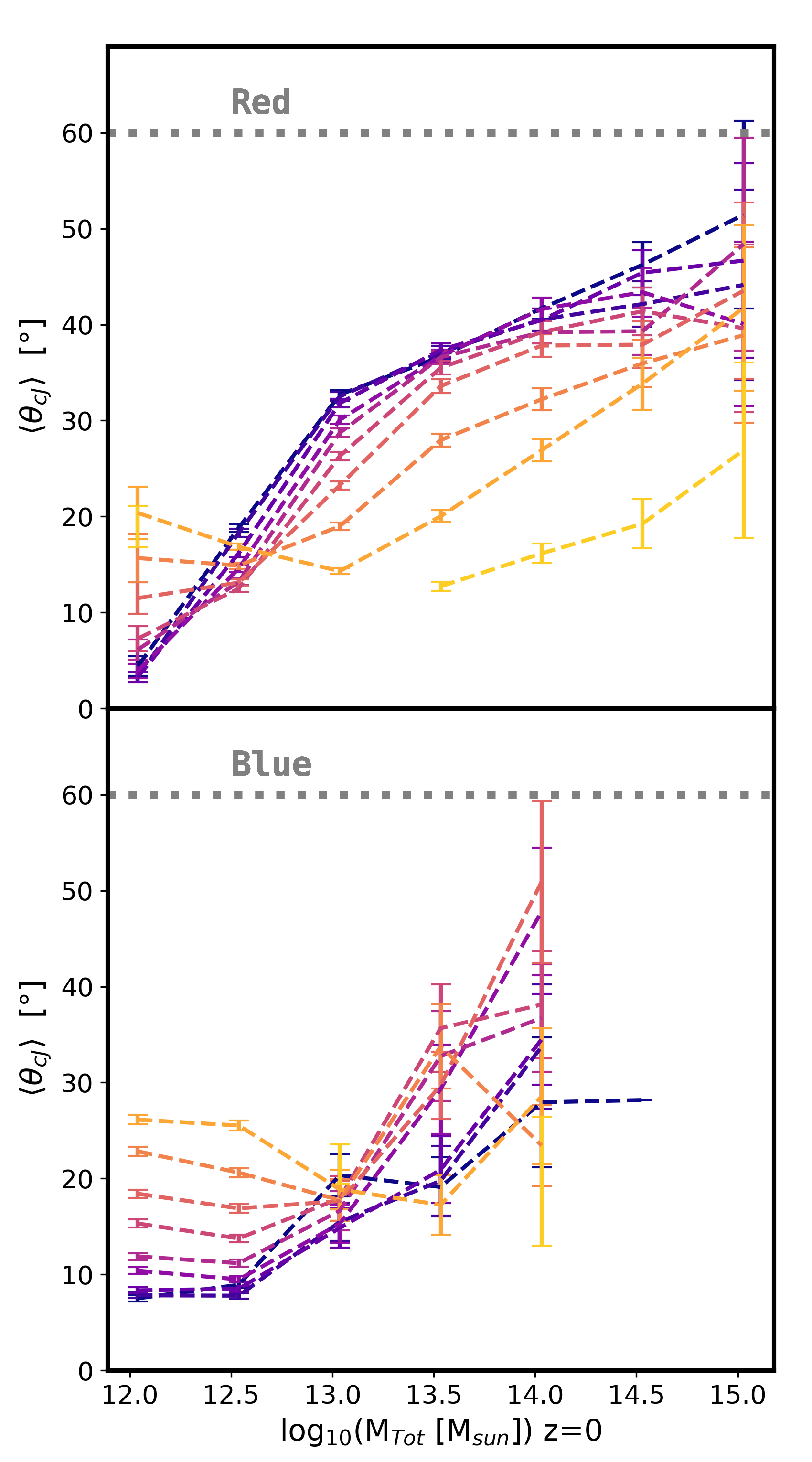}
    \caption{Evolution of the average misalignment angle  $\theta_{cJ}$  between the galaxy's angular momentum and  minor axis as a function of total mass. The upper panel shows the evolution of the alignments of  BGGs that are classified as red at $z=0$, i.e $(g-r)_{z=0}>0.6$, while the lower panel shows the evolution of  blue BGGs  i.e $(g-r)_{z=0}<0.6$.} 
   \label{cyL(masa)_color}
\end{figure}
\begin{figure}[ht!]
    \centering
    \includegraphics[width=0.4\textwidth]{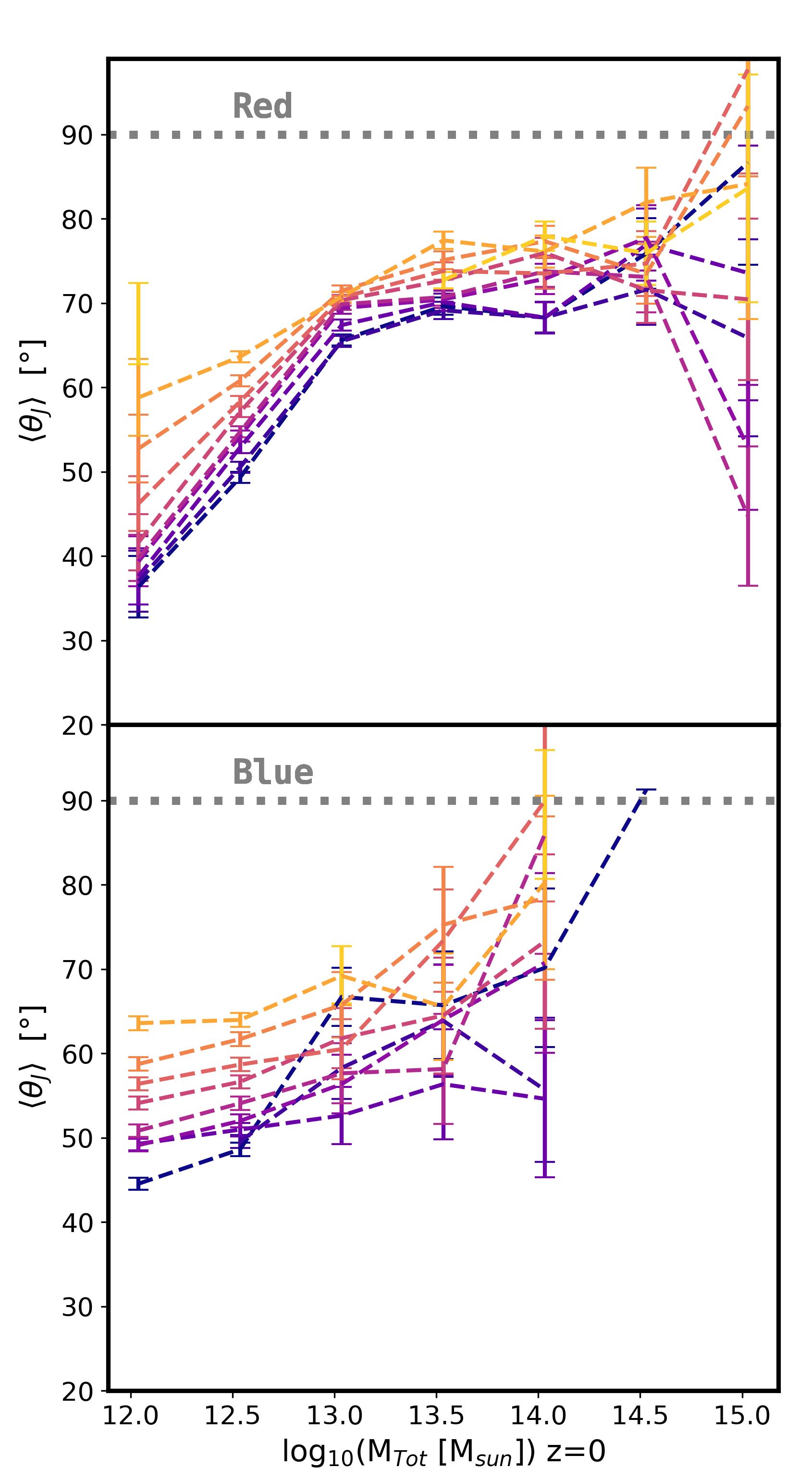}
    \caption{Evolution of the average misalignment angle  $\theta_J$  between the stellar and dark matter angular momenta as a function of total mass. The angle $ \theta_{J} $ ranges from $0$ to $180^\circ$, and for two random directions, the average value is $\langle \theta_J \rangle = 90^\circ $. Therefore the grey dotted line now indicates  an average angle of $90^\circ$, corresponding to random orientation, and smaller angles indicate a tendency toward alignment. The upper panel shows the evolution of the alignments of  BGGs that are classified as red at $z=0$, i.e $(g-r)_{z=0}>0.6$, while the lower panel shows the evolution of  blue BGGs  i.e $(g-r)_{z=0}<0.6$.} 
   \label{angL(masa)}
\end{figure}
\begin{figure*}
\sidecaption
  \includegraphics[width=12cm]{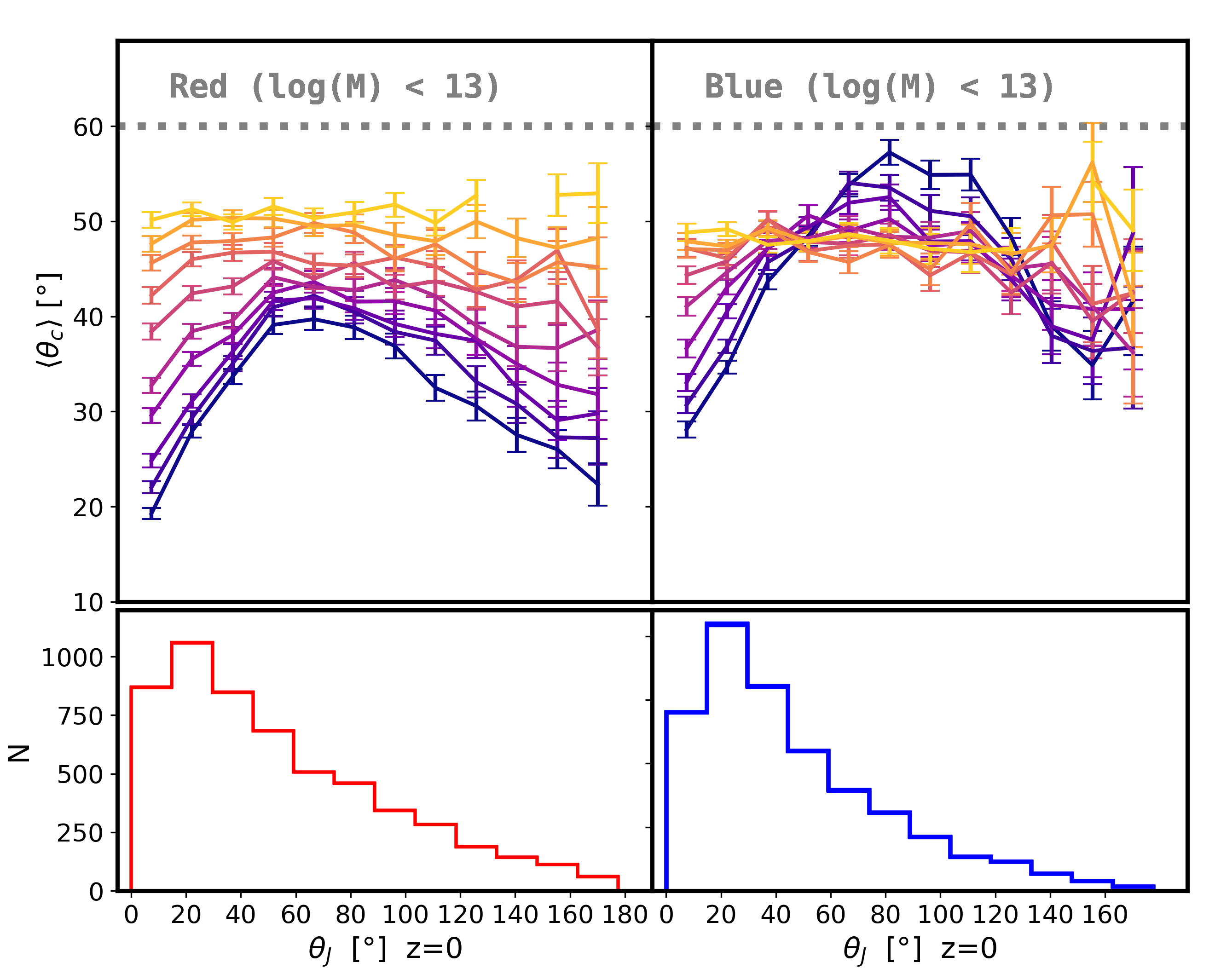}
     \caption{Upper panels show the evolution of the average misalignment angle  $\theta_c$  between the stellar and dark matter minor axis as a function of  their $\theta_{J} $ alignment at $z=0$, for red (left panel) and blue (right panel) BGGs within the mass range  $\mathrm{M_{Tot}} < 10^{13} \, \mathrm{M}_\odot$. The grey dotted line on the upper panels indicates  an average angle of $90^\circ$, corresponding to random orientation.  The lower panels illustrate the distribution of $\theta_{J} $ for red (left panel) and blue (right panel) BGGs within that same mass range.}
     \label{angC(angL)}
\end{figure*}

Figure \ref{cyL(masa)_color} shows the evolution of the angle $\theta{cJ}$, which measures the misalignment between a BGG’s minor axis and its stellar angular momentum, as a function of mass. In the upper panel, we see that red BGGs in the lower-mass regime ($\mathrm{M}_\mathrm{Tot} < 10^{13} \, \mathrm{M}_\odot$) have their minor axis tightly coupled to the stellar spin at $z=0$, although only those with $\mathrm{M}_\mathrm{Tot} \sim 10^{12} \, \mathrm{M}_\odot$ show a clear evolutionary trend toward stronger alignment. By contrast, red BGGs in higher-mass bins gradually lose this coupling, and for $\mathrm{M}_\mathrm{Tot} > 10^{13} \, \mathrm{M}_\odot$ the alignment steadily weakens, so that by $z=0$ the minor axis is only loosely connected to the stellar angular momentum. These massive galaxies likely inhabit denser environments and undergo more frequent interactions and mergers, which drive them toward dispersion-dominated stellar orbits. 
Moreover, several studies have shown that AGN activity is one of the main drivers of quenching in massive BGGs, both in IllustrisTNG \citep[e.g][]{Zinger2020,Donnari2020} and in other cosmological simulations such as Horizon-AGN \citep[e.g.,][]{Dubois2016,Beckmann2017}. Therefore, it would be interesting to further explore in future works the influence of  AGN feedback  on the observed alignment patterns in these systems. On the other hand, the lower panel shows that blue BGGs follow a different trend, displaying a tight $\theta_{cJ}$ alignment already at higher redshifts, which is further strengthened over time across most mass bins. The exception is the most massive systems ($\mathrm{M}_\mathrm{Tot} \sim 10^{13} \, \mathrm{M}_\odot$), which exhibit less evolution in this regard. By $z=0$, the lower-mass blue BGGs are oriented parallel to their angular momentum, consistent with the picture that at high redshift these galaxies were still assembling mass while, at later times, they settled into an equilibrium configuration where the disk is fully formed and its orientation is governed by the spin axis.

Therefore, the shape orientation of lower-mass BGGs appears to correlate with their stellar angular momentum at $z=0$, but the extent of its coupling to the dark matter angular momentum remains uncertain. 
Consequently, it is important to examine the connection between the dark matter and stellar spin axes. In Fig. \ref{angL(masa)}, we illustrate the evolution of the alignment angle $\theta_J$, which quantifies the orientation of the stellar angular momentum relative to that of the inner halo, for the blue and red populations in the lower and upper panels, respectively, as a function of mass. For both red and blue BGGs, we observe that from $z=2$ to $z=0$, the alignment improves across the entire mass range, although with a noisier trend at the high-mass end for blue BGGs.

Red BGGs exhibit a clear mass dependence, as the alignment angle $\theta_J$ shows a steep decrease with increasing mass for BGGs with $\mathrm{M}_\mathrm{Tot} < 10^{13} \, \mathrm{M}_\odot$. In contrast, higher-mass red BGGs display a much less pronounced evolution in $\theta_J$, largely independent of mass, so that by $z=0$ their stellar angular momentum is only weakly coupled to that of the dark matter. As we showed previously, the minor axis of these massive red BGGs is also not strongly aligned with the stellar angular momentum direction. In consequence, the angle $\theta_c$ for higher-mass red BGGs is unlikely to be significantly affected by large misalignments in $\theta_J$.

The alignment $\theta_J$ of blue BGGs evolves in a similar way, although with a slightly greater alignment leading to a  weaker mass dependence. Notably, at the low-mass end ($\mathrm{M}_\mathrm{Tot} < 10^{13} \, \mathrm{M}_\odot$), red BGGs exhibit $\theta_J$ alignment values of similar magnitude to those of blue BGGs. As we also showed earlier, these lower-mass  BGGs have a tighter coupling between their minor axis and the orientation of the stellar angular momentum. Therefore, it is important to understand the impact of the dynamical coupling between the stellar and dark matter angular momentum $\theta_J$ and the internal  alignments of red and blue BGGs in the low mass range.

To assess the direct influence of $\theta_J$ alignment on the galaxy–halo shape alignment, the upper panels of Fig. \ref{angC(angL)} present the evolution of $\theta_c$ from $z=2$ to $z=0$ as a function of the present-day ($z=0$) $\theta_J$ values, for red and blue BGGs in the left and right panels, respectively, within the lower-mass range ($\mathrm{M}_\mathrm{Tot} < 10^{13} \, \mathrm{M}\odot$).
At $z=2$, red BGGs exhibit similar $\theta_c$ alignment values across the full range of $\theta_J$. As redshift decreases, those BGGs whose stellar and dark matter angular momenta are aligned (either parallel or anti-parallel) at $z=0$ also tend to develop a strong alignment of their stellar and dark matter minor axes over time. Conversely, red BGGs whose present-day $\theta_J$ is close to perpendicular initially display a marked increase in $\theta_c$ alignment, but for $z \leq 0.6$, this evolution slows appreciably. Consequently, these galaxies do not achieve as tight an alignment between their stellar and dark matter shapes by $z=0$.

Blue BGGs at $z=2$ exhibit $\theta_c$ alignment values that are very similar to those of red BGGs; however, their subsequent evolution differs markedly. Among blue BGGs whose stellar and dark matter angular momenta are aligned (parallel or anti-parallel) at $z=0$, the galaxy–halo shape alignment strengthens as redshift decreases, although not to the same extent as observed for red BGGs. In contrast, for blue BGGs whose present-day $\theta_J$ is close to perpendicular, the $\theta_c$ alignment declines over time, such that by $z=0$ the orientation of their stellar component becomes nearly random.

Consequently, we observe that for lower-mass blue BGGs, the internal shape alignment is strongly coupled to the alignment between their stellar and dark matter angular momenta, as both $\theta_c$ and $\theta_J$ either increase or decrease together over time. Hence, differences in the evolutionary processes that blue BGGs experience are reflected in their present-day shape alignment. 
Moreover, we find that for blue BGGs, even small misalignment angles in $\theta_J$ are associated with larger shape misalignments ($\theta_c$) compared to red BGGs.
In contrast, lower-mass red BGGs always tend to align over time and are less sensitive to misalignments in angular momentum.

Therefore, even though Fig. \ref{angL(masa)}  shows that red and blue BGGs in the lower-mass range have similar $\theta_J$ alignments, the greater sensitivity of blue BGGs’ shape orientation even to small misalignments in $\theta_J$ explains why their present-day $\theta_c$ alignment is weaker. One possible mechanism that could produce these misalignments in blue BGGs, as suggested by \cite{Debattista2015}, is the accretion of gas whose angular momentum is misaligned with the halo, which helps to keep the stellar disk misaligned in shape relative to the dark matter component.
We also investigated possible dependencies of $\theta_J$ on the number of major mergers but did not find a clear correlation. Nonetheless, other merger-related processes might perturb the stellar disk orientation, and exploring these effects  would be an important direction for future work.

We also note  that the error bars in the upper panels of Fig. \ref{angC(angL)} increase with $\theta_J$, reflecting the decreasing number of galaxies per bin. This trend is evident in the lower panels of Fig. \ref{angC(angL)}, which show the $\theta_J$ distributions for red and blue BGGs in the left and right panels, respectively. Both distributions are very similar within this mass range: most galaxies occupy the parallel bins, fewer fall into the perpendicular bins, and only a small fraction lie in the antiparallel bins. As a result, it is relatively uncommon, particularly for blue BGGs, to rotate in the opposite direction to their halo.

\section{Summary and conclusions}  
\label{sec:conclusions}

In this paper, we investigate the evolution of the internal alignment between the principal axes of the stellar and dark matter components of BGGs identified at $z=0$ in the IllustrisTNG300-1 simulation, exploring potential drivers behind the contrasting alignment behavior observed for red and blue BGGs. This work is motivated by previous studies \cite{Rodriguez2023,Rodriguez2024}, which revealed a connection between the  alignment of BGGs on scales up to $\sim 10 \, \mathrm{Mpc}$ and their internal galaxy–halo shape alignment, with blue BGGs exhibiting a much weaker anisotropic correlation signal due to stronger internal misalignments.

First, we find that BGGs with greater total mass at $z=0$ exhibit the strongest increase in internal shape alignment as redshift decreases. In particular, central galaxies with $\mathrm{M_{Tot}} > 10^{13} \, \mathrm{M}_\odot$ show a marked enhancement in alignment, largely independent of other properties, although the most massive systems are predominantly red and have undergone multiple major mergers.
In this mass range, AGN feedback plays a fundamental role in quenching and in the emergence of the massive elliptical population \citep[e.g.,][]{Dubois2016,Beckmann2017,Donnari2020,Zinger2020}. Therefore, future studies should further investigate the impact of AGNs on internal alignments. 

Also,  massive BGGs that have experienced many major mergers tend to align their major axis with the halo's major axis earlier in their history, while those with fewer major mergers take longer to reach a similar alignment configuration.
This suggests that the process of mass accretion, along with the fact that these galaxies are generally dynamically younger systems, contributes to maintaining the connection between the halo's matter distribution and the stellar component’s shape.
This strong internal connection, together with the link between the outer  halo shape and the large scale anisotropies  \citep[e.g.,][]{Bailin2005,Valenzuela2024,Han2024,Rodriguez2025}, helps to explain  the reason for the larger anisotropy signal detected for these massive red galaxies, even in the two-halo term, in the works of \cite{Rodriguez2023,Rodriguez2024}.

Since the low-mass end is dominated by blue and late-type galaxies, which are typically oblate in shape, we focused on the direction of their minor axis, as it serves as the best tracer of their orientation. When exploring additional dependencies on color and the number of major mergers, we found that, among BGGs in the lower-mass regime ($\mathrm{M_{Tot}} < 10^{13} \, \mathrm{M}_\odot$), red   BGGs and those with a higher number of mergers exhibit the strongest evolution toward alignment, in terms of the $\theta_c$ angle.
However, given the correlation between galaxy color and merger history, we isolated the red population to assess the direct impact of mergers. This analysis revealed that color has a more dominant influence on the internal alignment than simply experiencing more mergers, although BGGs that have undergone multiple mergers still tend to have  more evolution towards alignment. 

In summary, major mergers contribute to quenching star formation and  BGGs with frequent mergers are more likely to be red and hence display stronger galaxy–halo alignments. Furthermore, once galaxies are quenched, subsequent mergers appear to reinforce this alignment.  
Therefore, it seems that once lower-mass galaxies lose their gas, cease star formation, and transition into red systems, their most stable configuration is one in which the minor axis of their stellar component aligns with the minor axis of the inner halo. In this context, as noted by \citep{Debattista2015}, ongoing gas inflows might play a key role in preventing blue BGGs from achieving a similar galaxy–halo alignment.

Given that red and blue galaxies usually present different morphological features, we verified that the alignment trends of early and late-type BGGS using the \(\kappa_{rot}\) parameter (see equation 1 in \citealt{Sales2012} and figure 13 in \citealt{Rodriguez2024}) are similar to, if not weaker than, those found when comparing red and blue BGGs, so we chose not to present them since they provide no additional insight.

To assess the influence of dynamical processes on the stellar–dark matter alignment of BGGs, we first tracked the evolution of their minor-axis orientation relative to their angular momentum, finding a clear mass dependence. 
At high redshift, present-day red and massive BGGs  typically had  their minor axis aligned with their angular momentum. However, as these systems underwent multiple complex baryonic processes, the angular momentum they retained gradually decoupled from their minor axis. Only red BGGs with $\mathrm{M_{Tot}} \leq 10^{12.5} \, \mathrm{M}_\odot$ show some increase in their minor axis–spin alignment over time. Interestingly, in the IllustrisTNG subgrid models, the kinetic mode of AGN feedback has an important effect on the galaxy's properties for stellar masses above $\sim10^{10.5} \, \mathrm{M}_\odot$, while at lower masses AGN activity occurs mainly through the thermal mode  \citep{Zinger2020}. This may be related to the mass scale at which we observe the transition in alignment behavior. Although,  the kinetic mode in IllustrisTNG is isotropic, so it is not clear  that it should induce any preferential direction, and further study is needed to determine if there is a correlation. In contrast, blue BGGs maintain a much tighter coupling between their minor axis and stellar angular momentum orientation across all redshifts and masses, highlighting the important role of tidal processes in shaping their internal alignment.

Furthermore, when considering the alignment between  the galaxy and  dark matter angular momenta $\theta_J$, we observed that, even though the alignment mildly increases  over time throughout the whole mass range, it is  stronger for lower mass BGGs too. 
Interestingly, red and blue BGGs in that mass range exhibit similar alignment values $\theta_J$ at $z=0$ but their evolution is different, with blue BGGs evolving towards alignment more strongly, while red BGGs only present a mild evolution. Consequently, it seems that red BGGs' $\theta_J$ alignment does not change much, but their $\theta_c$ alignment increases strongly over time, and blue BGGs present the opposite behavior; on average, they increase their $\theta_J$ alignment more strongly than their $\theta_c$ alignment.

Thus, we examined how the evolution of the internal alignment $\theta_c$ relates to the alignment angle $\theta_J$ fixed at $z=0$,  for red and blue BGGs with $\mathrm{M_{Tot}} < 10^{13} \, \mathrm{M}_\odot$. Our analysis shows that $\theta_c$ in blue BGGs is substantially more sensitive to the stellar–dark matter angular momentum alignment. Specifically, blue BGGs with nearly perpendicular $\theta_J$ values at $z=0$ tend to exhibit a decreasing $\theta_c$ alignment as redshift decreases, whereas those with nearly parallel (or anti-parallel) $\theta_J$ values tend to progressively align their minor axis with that of the inner halo.
In contrast, red BGGs consistently strengthen their galaxy–halo shape alignment over time, largely independent of $\theta_J$, although this growth is slower for those with nearly perpendicular $\theta_J$ values at $z=0$.

These results suggest that, for some blue BGGs whose dynamical state has not been significantly perturbed, their stellar angular momentum progressively aligns with the dark matter angular momentum orientation, and likewise, their minor axis aligns strongly with their halo’s minor axis. However, on average, blue BGGs present a mild  evolution towards alignment of $\theta_J$  and less so  for $\theta_c$.
On the other hand, those with nearly random $\theta_J$ and $\theta_c$ values at $z=0$ seem to have undergone violent processes that altered both their stellar dynamics and shape orientation. 

The findings of \cite{Debattista2015} point to the infall of gas with misaligned angular momentum as a primary mechanism for perturbing the stellar angular momentum orientation, thereby maintaining a misalignment between the inner halo and the disk. Then, this mechanism might give rise to these latter cases. Those blue BGGs with anti-parallel  $\theta_J$ also seem to be particular and rare systems that slightly align their minor axis with that of the halo but end up rotating in the opposite direction with respect to the halo's angular momentum. 

In the work of \cite{Romeo2022}, spectroscopic data of late-type galaxies were used to investigate the connection between their specific angular momentum and other key astrophysical properties. The authors found that the fraction of baryonic angular momentum relative to the angular momentum of the halo $\mathrm{j_b}/\mathrm{j_h}$ is larger for galaxies with higher baryonic mass fractions $\mathrm{M_b}/\mathrm{M_h}$, meaning that these latter tend to retain their angular momentum more effectively. 
They argue that this mass dependence reflects a complex interplay of baryonic processes, including galactic outflows driven by stellar and AGN feedback, as well as differences in the transfer of angular momentum from the gas to the halo in galaxies residing in more massive halos. Therefore, future studies could further investigate whether these mechanisms are connected to the alignment patterns we observe.

Ultimately, it might be the case that once low mass BGGs exhaust their gas supply and become quenched systems,  their stellar component  may gradually lose  angular momentum which remains more or less aligned with the halo's spin, but this does not prevent them from evolving towards alignment with the dark matter shape. Nevertheless, other processes  may also perturb red lower-mass BGGs, potentially explaining the existence of those that have nearly random $\theta_J$ and reduced minor-axis alignment  at $z=0$. Although these are not significant enough to reverse the tendency toward alignment.

Furthermore, blue BGGs with nearly random values of $\theta_J$ and $\theta_c$ at $z=0$ may be in a similar evolutionary stage to high-redshift BGGs that also exhibited large misalignments, but have since evolved into massive, red systems. These already perturbed disks might eventually become the red systems that, as we explained above, have their minor-axis orientation more strongly aligned with that of the halo, yet still retain a relatively large $\theta_J$ misalignment, due to the fact that their stellar shape is not as strongly governed by the stellar angular momentum orientation.

Consequently, our results indicate that blue BGGs consistently display weaker internal shape alignments across the entire mass range, driven by the various processes discussed in this work. This naturally accounts for the absence of a large-scale anisotropy signal in blue BGGs, as reported by \cite{Rodriguez2023,Rodriguez2024}.
Future investigations should aim to clarify the role of mergers in reinforcing BGG alignments beyond the quenching effect, for instance by examining correlations with merger-related properties other than the number of major mergers. Additional insights may be gained by exploring alignments involving the gaseous component and by tracing the detailed evolutionary histories of individual systems to examine what processes are responsible for the population of blue BGGs with misaligned $\theta_J$ and $\theta_c$.

Given that we build upon the analysis of the alignment signal previously investigated by \cite{Rodriguez2023,Rodriguez2024}, we retained the original magnitude cut  ($\mathrm{M_r} < -21.5$) adopted by  the authors for consistency. However, we also used a lower magnitude threshold ($\mathrm{M_r} < -19.5$) to test the robustness of our results. This broader sample includes a wider mass range at the low-mass end, a larger proportion of blue BGGs, and a significant shift in the distribution of major mergers toward lower merger counts. Nevertheless, the main conclusions regarding alignment trends remain valid for this fainter sample (see Appendix \ref{sec:Appendix}).

Moreover, our results are in qualitative agreement with the findings of \cite{Bhowmick2019}, who tracked the evolution of the galaxy–halo major axis alignment for different mass samples fixed at $z=0$ in the MassiveBlack-II cosmological simulation. They reported an increasing alignment trend with both decreasing redshift and increasing halo mass. On the other hand, \cite{Bate2019} used the Horizon-AGN simulation to study the alignment between the galaxy major axis and the tidal field minor axis for the progenitors of elliptical galaxies selected at $z=0$, finding comparable mass and redshift dependencies. Although this latter study examines a different type of alignment, the consistent mass and redshift trends suggest a possible connection between the galaxy–halo shape alignment analyzed here and alignments with the tidal field which could also be examined in future works.

Also, even though it is not possible to study these internal alignments directly in observations, forthcoming spectroscopic surveys such as Euclid and DESI will provide larger and deeper datasets, and they will open up an opportunity  to explore  with other approaches, such as lensing, whether observations are consistent with our findings. At the same time, probing the evolution  of alignments observationally also remains challenging, but may become feasible by using, for example, Ly$\alpha$-emitting galaxies (LAEs) detected at high redshift in the ODIN survey \citep{Lee2024ODINsurvey} as potential progenitors of present-day BGGs. In this context, a deeper understanding of galaxy–halo alignments holds the potential to transform our view of how galaxies inherit, retain, or lose the imprint of their dark matter halos.

\begin{acknowledgements}
    The authors wish to thank the anonymous referee for her/his report that helped us to improve this manuscript. AVMC, FR and  MM  thank the support by Agencia Nacional de Promoci\'on Cient\'ifica y Tecnol\'ogica, the Consejo Nacional de Investigaciones Cient\'{\i}ficas y T\'ecnicas (CONICET, Argentina) and the Secretar\'{\i}a de Ciencia y Tecnolog\'{\i}a de la Universidad Nacional de C\'ordoba (SeCyT-UNC, Argentina).
    FR acknowledges support from the ICTP through the Junior Associates Programme 2023-2028. 
\end{acknowledgements}

\bibliographystyle{aa} 
\bibliography{bibliografia.bib}

\begin{appendix} 
\section{Extended sample with $\mathrm{M_r} < -19.5$}
\label{sec:Appendix}

To assess the robustness of our results, we repeated the analysis using a broader BGG sample, adopting a fainter magnitude limit of $\mathrm{M_r} < -19.5$ instead of $\mathrm{M_r} < -21.5$. This naturally adds more lower-mass galaxies, extending the lower bound of the mass range to $\sim10^{11} \, \mathrm{M_\odot}$. However, shape determinations for these systems at high redshift may be less reliable due to the smaller number of star particles tracing their structure. In this broader sample, blue BGGs outnumber red ones, and the major-merger distribution shifts toward lower counts, with most BGGs experiencing fewer than two mergers with a stellar mass ratio of 0.2. 

\begin{figure}[ht!]
    \centering
    \includegraphics[width=0.4\textwidth]{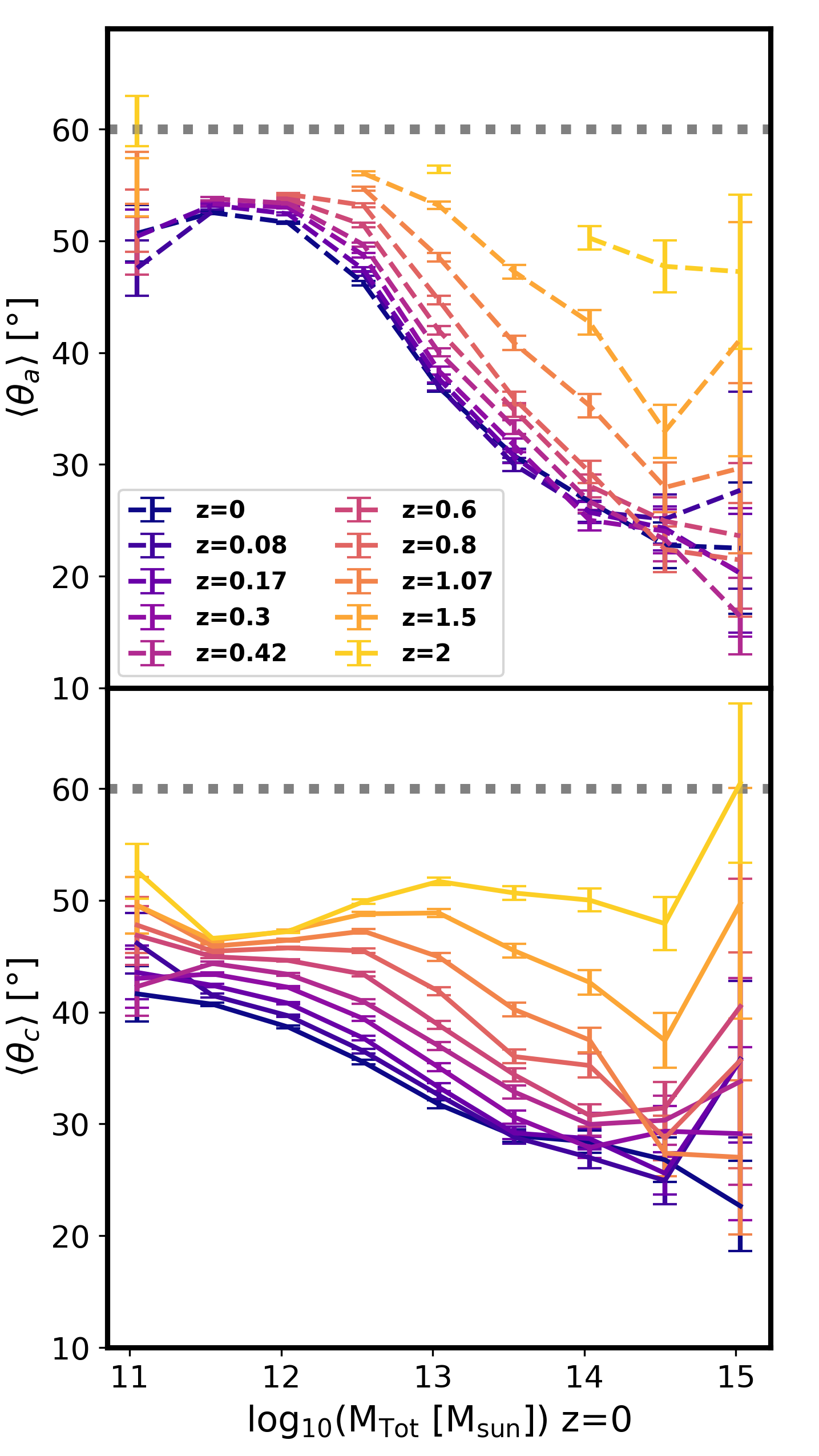}
     \caption{Average misalignment angles  $\theta_a$ (upper panel) and $\theta_c$ (lower panel) of the  stellar and dark matter major and minor axes respectively calculated at redshifts $z=0,0.8,0.17,0.3,0.42,0.6,1.07,1.5,2$, as a function of the total mass at $z=0$,  for BGGs with absolute magnitude  in the r-band $\mathrm{M}_r < -19.5$. The grey dotted line indicates an average angle of $60^{\circ}$, corresponding to random orientation, smaller angles indicate a tendency toward alignment. Error bars are calculated using the standard deviation of the mean.}
     \label{ayc(mass)-195}
\end{figure}
\begin{figure}[ht!]
    \centering
    \includegraphics[width=0.4\textwidth]{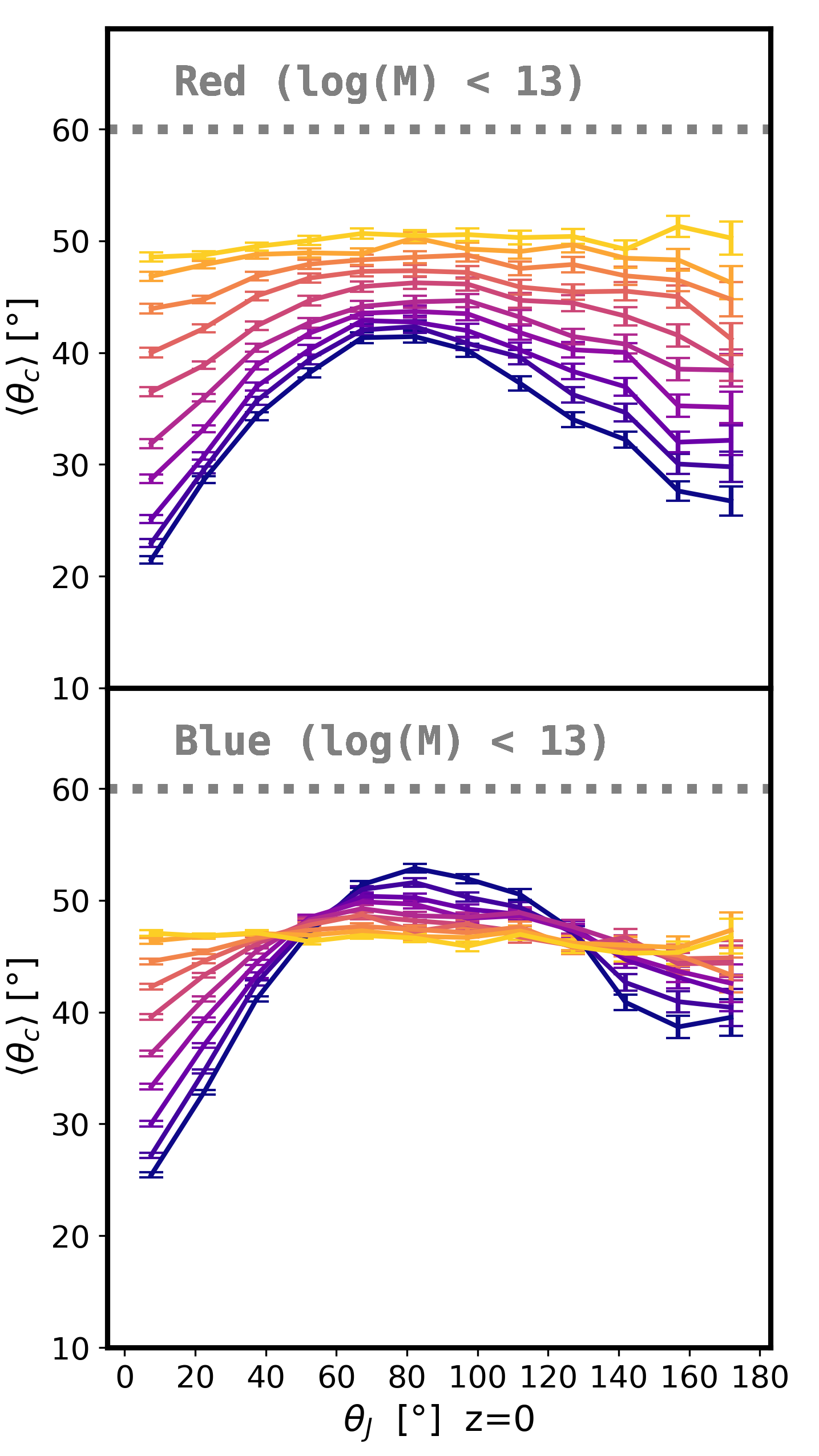}
    \caption{Evolution of the average misalignment angle  $\theta_c$  between the stellar and dark matter minor axis as a function of their alignment $\theta_{J} $ at $z=0$, for BGGs within the mass range  $\mathrm{M}_\mathrm{Tot} < 10^{13} \, \mathrm{M}_\odot$ and with absolute magnitude  in the r-band $\mathrm{M}_r < -19.5$.  The upper panel shows the evolution of the alignments of  BGGs that are classified as red at $z=0$, i.e $(g-r)_{z=0}>0.6$, while the lower panel shows the evolution of  blue BGGs  i.e $(g-r)_{z=0}<0.6$. } 
   \label{angC(angL)-195}
\end{figure}
In Fig. \ref{ayc(mass)-195}, we show the major and minor axis alignments for the total sample in the upper and lower panels, respectively, as a function of total mass.
For the fainter sample, two additional low-mass bins appear, exhibiting trends similar to those of galaxies with
$ \mathrm{M_{Tot}} \sim 10^{12} \, \mathrm{M_\odot} $ in the lowest-mass bin of the brighter sample, but with smaller error bars owing to the larger number of BGGs.
At the high-mass end, the alignment evolution remains essentially unchanged compared to the brighter sample.

As before, we find that for BGGs with $ \mathrm{M_{Tot}} < 10^{13} , \mathrm{M_\odot} $, red systems strengthen their $\theta_c$ alignment more noticeably than blue counterparts of the same mass as redshift decreases in this extended sample.
For comparison, we illustrate in Fig. \ref{angC(angL)-195} the relation between $\theta_c$ and $\theta_J$ at $z=0$ for red and blue BGGs in the upper and lower panels, respectively, restricted to $ \mathrm{M_{Tot}} < 10^{13} , \mathrm{M_\odot} $.
The lower-mass galaxies in this fainter sample follow the same trends observed in the bright-galaxy sample. Specifically, blue BGGs display a galaxy–halo alignment that is more closely coupled to the stellar–dark matter angular momentum alignment, with both angles increasing or decreasing together over time. In contrast, red BGGs consistently increase their internal shape alignment, although this growth is more modest for red discs with misaligned angular momentum.

\end{appendix}

\end{document}